\documentclass[sigconf]{acmart}

\usepackage{enumitem} 
\usepackage{multirow}
\usepackage{graphicx}
\usepackage{subcaption}
\usepackage{booktabs}
\usepackage{soul}
\usepackage{adjustbox}
\usepackage{xcolor}
\usepackage{balance}

\DeclareMathAlphabet{\pazocal}{OMS}{zplm}{m}{n}
\DeclareMathAlphabet{\pazobfcal}{OMS}{cmsy}{b}{n}
\DeclareMathOperator*{\avg}{avg}
\DeclareMathOperator*{\G}{G}

\newcommand{\uls}{\begin{itemize}[leftmargin=*]}
\newcommand{\ule}{\end{itemize}}
\newcommand{\ols}{\begin{enumerate}[leftmargin=*]}
\newcommand{\ole}{\end{enumerate}}
\newcommand{\li}{\item}
\newcommand{\para}[1]{\paragraph{\textnormal{\textbf{#1}}}}

\AtBeginDocument{%
}

\copyrightyear{2025}
\acmYear{2025}
\setcopyright{cc}
\setcctype{by}
\acmConference[CIKM '25]{Proceedings of the 34th ACM International Conference on Information and Knowledge Management}{November 10--14, 2025}{Seoul, Republic of Korea}
\acmBooktitle{Proceedings of the 34th ACM International Conference on Information and Knowledge Management (CIKM '25), November 10--14, 2025, Seoul, Republic of Korea}
\acmDOI{10.1145/3746252.3760820}
\acmISBN{979-8-4007-2040-6/2025/11}

\begin{document}

\title{T-Retrievability: A Topic-Focused Approach to Measure Fair Document Exposure in Information Retrieval}

\author{Xuejun Chang}
\affiliation{
  \institution{University of Glasgow}
  \city{Glasgow}
  \country{UK}
}
\email{x.chang.2@research.gla.ac.uk}

\author{Zaiqiao Meng}
\affiliation{
  \institution{University of Glasgow}
  \city{Glasgow}
  \country{UK}
}
\email{Zaiqiao.Meng@glasgow.ac.uk}

\author{Debasis Ganguly}
\affiliation{
  \institution{University of Glasgow}
  \city{Glasgow}
  \country{UK}
}
\email{Debasis.Ganguly@glasgow.ac.uk}

\renewcommand{\shortauthors}{Xuejun Chang, Zaiqiao Meng, \& Debasis Ganguly}

\begin{abstract}    
Retrievability of a document is a collection-based statistic that measures its expected (reciprocal) rank of being retrieved within a specific rank cut-off. A collection with uniformly distributed retrievability scores across documents is an indicator of fair document exposure. While retrievability scores have been used to quantify the fairness of exposure for a collection, in our work, we use the distribution of retrievability scores to measure the exposure bias of retrieval models. We hypothesise that an uneven distribution of retrievability scores across the entire collection may not accurately reflect exposure bias but rather indicate variations in topical relevance. As a solution, we propose a topic-focused localised retrievability measure, which we call \textit{T-Retrievability} (topic-retrievability), which first computes retrievability scores over multiple groups of topically-related documents, and then aggregates these localised values to obtain the collection-level statistics. Our analysis using this proposed T-Retrievability measure uncovers new insights into the exposure characteristics of various neural ranking models. The findings suggest that this localised measure provides a more nuanced understanding of exposure fairness, 
\end{abstract}


\maketitle

\section{Introduction}

Supervised information retrieval (IR) models based on pre-trained transformer architectures, such as ColBERT \cite{colbert} and Mono-T5 \cite{monot5}, have demonstrated superior relevance ranking performance compared to their unsupervised counterparts, such as BM25 \cite{BM25_RobertsonWJHG94}. However, this paradigm shift necessitates the consideration of additional dimensions of effectiveness, including those of efficiency \cite{BruchNRV24,chang2024neural}, explainability \cite{VermaG19,DBLP:conf/ecir/PandianGM24,DBLP:conf/sigir/SenGVJ20}, and exposure fairness \cite{singh2018fairness,FairnessExposure2022,patro2022fair}. In this study, we specifically examine the exposure fairness criterion of search systems, which has significant societal implications. For instance, ensuring a balanced distribution of topically relevant documents from diverse sources and ideological viewpoints is essential for preserving the integrity of information ecosystems \cite{patro2022fair}, fostering an informed and unbiased public opinion, and supporting democratic decision-making \cite{opinion,krause2024relevance,abolghasemi2024measuring}.

The data-driven learning of relevance in neural models makes it challenging to identify and mitigate latent biases present in the training data, which can manifest as algorithmic preferences leading to exposure biases \cite{patro2022fair,Mitigating_Popularity_Bias2023,musto2021fairness}. These exposure biases are expected ranks of retrieved documents computed over a large set of queries that several users may execute on a collection.
In the case of queries with similar information needs and overlapping sets of relevant documents, an IR model that aims to ensure exposure fairness should distribute the rankings of these documents more evenly by shuffling their positions across queries \cite{Fernando_2020,Ekstrand2023}. Assessing whether IR models achieve this fairness objective is crucial for understanding and addressing potential biases in retrieval systems.
We restrict the scope of our investigation of exposure bias to only document collections without categorical meta-data attributes, e.g., gender, polarity etc., as studied in \cite{rekabsaz2021societal}.

Previous research has introduced the concept of \textit{retrievability} as a measure of document accessibility \cite{leifRetrievabilityAnevaluationMeasure,retr-analysis-cikm2014,wilkie2015retrievability,Topical_Retrievability2016}. This collection-based statistic quantifies a document’s expected (reciprocal) rank of retrieval within a predefined rank cut-off, averaged over a sufficiently large and topically diverse set of queries. While retrievability analysis has traditionally been used to assess document accessibility across different collections \cite{retr-analysis-cikm2014}, more recent studies have applied it to examine exposure bias across various IR models \cite{synth-queries}.
However, two key considerations must be taken into account when using retrievability for such analyses. First, most existing studies, including those evaluating exposure fairness in IR models \cite{synth-queries}, rely on synthetic queries generated via word $n$-gram sampling. Synthetic queries used in prior research predominantly consist of unigrams or bigrams, whereas real-world user queries tend to be longer on average. Second, non-uniform retrievability values often arise due to variations in relevance priors, where certain high-quality documents are inherently more likely to be relevant than others. In such cases, disparities in retrievability should be accounted for by aggregating locally computed retrievability measures over groups of similar information needs.

The contributions of this paper are two-fold: a) addressing the limitation of synthetic queries by conducting retrievability analysis on a large set of real queries (MS MARCO dev set), aligning with real-world information needs; and b) proposing a topic-focused localized retrievability measure, named \textit{T-Retrievability} (topical retrievability), to conduct more fine-grained analysis of document accessibility of various ranking models.

\section{Proposed Measure for Exposure Fairness} \label{sec:method}

\para{Background on retrievability}
As per the originally proposed definition of retrievability \cite{leifRetrievabilityAnevaluationMeasure}, its value for a document $D \in \pazocal{C}$ (a collection of documents) indicates the likelihood of the document to be retrieved within a specified cut-off rank. As such, the retrievability value of a document depends on the retrieval model used to induce the ranking itself, and a cut-off rank is usually set to a small number, such as 10, indicating the size of a typical search results page \cite{serpsize}. More formally,
$r(D, \pazocal{C}, \pazocal{Q},\theta, k) = \frac{1}{|\pazocal{Q}|}\sum_{Q \in \pazocal{Q}} \mathbb{I}[\rho(D;Q, \theta) \leq k]$,
where $D$ is a document of the collection $\pazocal{C}$, $\pazocal{Q}$ denotes a set of queries, $\theta: D \times Q \mapsto \mathbb{R}$ denotes an IR model that assigns scores to documents thereby inducing a rank for $D$ denoted as $\rho(D;Q, \theta)$, which is a function of both a particular query $Q$ and the IR model $\theta$. The parameter $k$ denotes a cut-off rank and $\mathbb{I}[X]$ denotes an indicator function which is 1 if the condition $X$ is true, otherwise 0.

\para{Removing dependence on rank cut-off}
In this paper, we work with a slightly modified version of the retrievability measure,
where we remove the dependence of the cut-off rank $k$. More specifically, instead of computing the likelihood of a document to be retrieved within a top-$k$, we rather compute the expected rank of a document \cite{sinha2023findability}. In particular, to interpret retrievability as a score, i.e., the higher the better, we work with reciprocals of ranks instead of the ranks themselves. Formally,
\begin{equation}
r(D, \pazocal{C}, \pazocal{Q},\theta) = \frac{1}{|\pazocal{Q}|}\sum_{Q \in \pazocal{Q}} \frac{1}{\log(1+\rho(D;Q, \theta))} \label{eq:our-ret},    
\end{equation}
As a practical consideration, we restrict the expected rank computation to the top-100 set for each query. The advantage of Equation \ref{eq:our-ret} is that it is a rank-based measure, in contrast to being a set-based measure within top-$k$, and hence is also less likely to be sensitive to particular choices of the rank cut-off (100 in our case).

The bias in information access is computed by the Gini coefficient \cite{gini1921measurement} over the retrievability distribution of each document in the collection - a high value of the coefficient indicating a high disparity between document accessibility.

\para{From simulated to human-formulated queries}
An adequately large set of representative queries $\pazocal{Q}$ is used to measure the accessibility of a document in a collection $\pazocal{C}$ (Equation \ref{eq:our-ret}). All previous work on retrievability analysis \cite{leifRetrievabilityAnevaluationMeasure,retr-analysis-cikm2014,Retrievability_Azzo2015,Topical_Retrievability2016,retrievability_Ganguly2016} uses simulation to create this set $\pazocal{Q}$ - specifically via sampling words and their bi-grams from the collection. The $r(D)$ values can thus be affected by sampling biases resulting from the simulation. Instead, in our work, we conduct our experiments on real queries formulated by humans. Specifically, we employ the MS MARCO dev set as $\pazocal{Q}$ comprising a total of over 100K queries (collected from the Bing search log \cite{bajaj2016ms}).

The retrievability distribution thus computed over a large set of real queries for a number of different IR models reflects a more realistic information need. Furthermore, another reason to employ real queries is that it makes it possible to analyse the effects of document accessibility of neural rankers on the same dataset on which they are actually trained.

A likely reason why real-world query logs are often avoided in retrievability studies \cite{leifRetrievabilityAnevaluationMeasure,retr-analysis-cikm2014,Retrievability_Azzo2015,Topical_Retrievability2016,retrievability_Ganguly2016} is that they tend to be biased toward specific topics or user interests. However, employing a sufficiently large number of such human-formulated queries from search logs, as in our work, can mitigate this bias, as the topical skew diminishes when aggregated across a broader query set.

\para{Problem with non-uniform relevance-priors}
Previous research has shown that the prior distribution of relevance in a collection is usually non-uniform \cite{chang2024neural}, meaning that some documents due to their inherent characteristics, such as their lengths \cite{doclength-pivot}, citation counts \cite{docprior-pagerank} etc., are more likely to be relevant for a higher number of queries than others.
However, this means that this non-uniformity in relevance-priors is also likely to be manifested as non-uniformities in the collection-level retrievability scores (see Equation \ref{eq:our-ret}) computed for a retrieval model that effectively models relevance, thereby falsely penalising the model. 

\para{Localized Retrievability}
To address the influence of relevance priors on a collection-level document accessibility measure, such as retrievability, we propose to measure document accessibility for groups of topically related queries. This idea aligns with the per-query analysis of retrievability over a pool of retrieved documents as studied in \cite{Topical_Retrievability2016}.
This idea also aligns with the existing research on measuring exposure fairness for groups defined by document metadata (attribute value) combinations, such as demographics \cite{rekabsaz2021societal}, location \cite{Ekstrand2023}, and polarities \cite{opinion}. In our setup, instead of measuring disparities across groups of attribute values, we investigate disparities in document accessibility.

\para{Grouping the Queries}
For a fine-grained computation of retrievability scores, we partition the query set $\pazocal{Q}$ (in our experiments, the MS MARCO development queries) into $K$ clusters \cite{msmarco-clusters}. Specifically, we apply K-means clustering using two types of document vector representations: sparse (lexical) embeddings and dense (semantic) embeddings. These representations yield query groupings with distinct characteristics—one emphasizing precision and the other recall. The lexical approach is more conservative, grouping queries into the same cluster only when they share exact term matches, thereby achieving higher precision. In contrast, the semantic approach expands clusters based on term-level semantics, which increases recall but introduces a greater risk of reduced precision within each cluster.

For the sparse representation, we employ the standard TF–IDF vectorization of each query. For the dense representation, we use the [CLS] token's vector computed with the Sentence-BERT (SBERT) model \cite{reimers-gurevych-2019-sentence}. To ensure fair comparison across neural models, we adopt the \texttt{all-MiniLM-L6-v2} variant of SBERT -- a model pre-trained to capture semantic similarities \cite{bowman-etal-2015-large}.

\para{From localised measures to a collection-level measure}

On the partitioned query set $\pazocal{Q} = \cup_{i=1}^K \pazocal{Q}_i$, we then obtain $K$ such localised T-Retrievability scores - one for each group, i.e.,
\begin{equation}
r(D, \pazocal{C}, \pazocal{Q}_i,\theta) = \frac{1}{|\pazocal{Q}_i|}\sum_{Q \in \pazocal{Q}_i} \frac{1}{\log(1+\rho(D;Q, \theta))
}.
\label{eq:loc-retr}
\end{equation}
After computing the localised retrievability measures for each topic group (as shown in Equation \ref{eq:loc-retr}), the next step is to compute the document exposure fairness for each of these topic groups by computing $K$ different Gini coefficients. These can then be aggregated (minimum, maximum, or average) to obtain an overall document accessibility measure for a specific retrieval model $\theta$ on a collection $\pazocal{C}$.
Each way of aggregating corresponds to an analysis of the best case (minimum), average case, or worst case (maximum) exposure fairness of documents.
More formally speaking,
\begin{equation}
\bar{G}[r(D, \pazocal{C}, \pazocal{Q},\theta),\oplus] = \oplus_{i=1}^K \G[r(D, \pazocal{C}, \pazocal{Q}_i,\theta)],
\label{eq:locret-aggr}
\end{equation}
where $\bar{G}[\cdot]$ represents the Gini
coefficient of a set of localised retrievability values, $\oplus \in \{\min, \avg,\max\}$ denotes an aggregation operator.
We call this new collection-level measure of document accessibility (Equation \ref{eq:locret-aggr}) by the name \textbf{Topical-Retrievability}, abbreviated as \textbf{T-Retrievability}\footnote{The implementation is available at \url{https://github.com/XuejunChang/T-Retrievability}
}.

\begin{table}[t]
\centering
\small
\caption{A comparative analysis of the exposure fairness of different retrieval models on MS MARCO dev set. `G' denotes the Gini coefficient computed over the collection-level retrievability scores, whereas $\G_\oplus$ denotes an aggregation ($\oplus \in \{\min, \avg, \max\}$) of the topical retrievability values (see Equations \ref{eq:loc-retr} and \ref{eq:locret-aggr}), calculated by K-means clustering with dense vector representations of the dev set. The number of groups ($K$) for the queries of the dev set is set to 5000. 
\label{tab:exposure-bias-results}
}
\begin{tabular}{@{}l c c c c c c@{}}
\toprule
& \multicolumn{4}{c}{Exposure fairness$\downarrow$} & \multicolumn{2}{c}{Relevance$\uparrow$} \\ 
\cmidrule(r){2-5} \cmidrule(r){6-7}
IR Model & $\G$ & $\G_{\min}$ & $\G_{\avg}$ & $\G_{\max}$ & nDCG@10 & MAP@100 \\
\midrule
BM25 & .4731	          & .1843 &	.2878         & .7412          & .2131 & .1781 \\
SPLADE & \textbf{.3948}   & .1843 & .3057         & .5799 & \textbf{.4460} & \textbf{.3863} \\
TCT & .3994       & .1843 &	.2970         & \textbf{.5417}          & .4210 & .3648 \\
BM25>>TCT &	.4473 &	.1843 &	.2832         & .7202          & .3713 & .3226 \\
Mono-T5 &	.4428     &	\.1843 &\textbf{.2818} & .7122          & .3962 & .3469 \\
\bottomrule
\end{tabular}
\end{table}

\section{Experiment Setup} \label{sec:expt}
The objective of our experiments is to compare the exposure biases of various types of ranking models, and also explore how these biases relate to the relevance-based effectiveness measures of these models. Specifically, our first research question is:
\uls
\li \textbf{RQ-1}: How sensitive is exposure bias to a ranking model?
\ule
Next, regarding measuring exposure bias, we investigate whether our proposed fine-grained approach with different aggregating mechanisms and the collection-level measure are correlated with each other, or they yield different system preferences in terms of exposure fairness. Explicitly,
\uls
\li \textbf{RQ-2}: How correlated is the topic-based exposure bias measure with the collection-level one? 
\ule
It is worth noting that our proposed localized retrievability scores depend on the granularity of the topical relatedness of queries, i.e., too coarse topics may lead to relevance prior biases, as is expected to be the case with the standard collection-level retrievability \cite{leifRetrievabilityAnevaluationMeasure}, whereas too fine-grained topics may fail to capture a collection-level accessibility \cite{Topical_Retrievability2016}. Our third research question is thus:
\uls
\li \textbf{RQ-3}: How sensitive is the proposed T-Retrievability measure on the granularity of the query groups?
\ule
Since it is possible that the query grouping itself (conservative grouping by lexical representation vs. more aggressive grouping by a semantic approach) may lead to different trends in exposure fairness as computed by the T-Retrievability measure, our next research question thus explores the sensitivity of the T-Retrievability measure to the query representation used to construct the topics.
\uls
\li \textbf{RQ-4}: How sensitive is the T-Retrievability measure on different embeddings (sparse vs. dense) used to cluster the queries?
\ule 

\para{Dataset}
All our experiments corresponding to retrievability analysis and evaluating the relevance of different rankers are conducted based on the MS MARCO dev dataset, which is comprised of 101,093 queries and their associated relevant documents.
The underlying collection of documents used is the MS MARCO passage collection, comprised of over 8.8 million passages~\cite{bajaj2016ms}.

Although it is common practice to evaluate the effectiveness of IR models using standard test collections with depth-pooled relevance assessments (e.g., DL, Robust), such collections are not suitable for assessing exposure fairness. By definition, retrievability is the expected (reciprocal) rank of documents across queries (Equation \ref{eq:our-ret}). According to the law of large numbers, this expectation is meaningful only when computed over a sufficiently large set of queries, which small benchmark collections cannot provide.

\para{IR Models}
To investigate what characteristic differences may cause variations in document accessibility, we experiment with different classes of retrieval models - namely, sparse, learned-sparse, sparse with reranking, and dense end-to-end, as listed below.
\uls
\li \textbf{BM25} 
\cite{BM25_RobertsonWJHG94}: BM25 is a member of the \textit{sparse} family of IR models; the term weighting scheme is a combination of the relative importance of terms within documents, term informativeness and document lengths.
\li \textbf{SPLADE} \cite{SPLADE2021}: SPLADE, which belongs to the \textit{learned-sparse} family of rankers, combines the posterior likelihoods of the masked language model (MLM) objective of BERT, and the noise contrastive estimation likelihoods to derive the token weights. 
\li \textbf{TCT-ColBERT} (abbreviated in Table \ref{tab:exposure-bias-results} as `\textbf{TCT}') \cite{tctcolbert2021}: This is a \textit{dense end-to-end} model distilled from ColBERT that learns effective document encoded vectors. An approximate nearest neighbour search on these embedded representations has been reported to yield better performance than the [CLS] pooling or mean pooling of the ColBERT model. In our experiments, we use the \texttt{pyterrier\_dr} library to construct the dense index.
\li \textbf{BM25>>TCT-ColBERT} (abbreviated in Table \ref{tab:exposure-bias-results} as `\textbf{BM25>>TCT}') \cite{tctcolbert2021}: the \textit{retrieve-and-rerank} version of TCT-ColBERT, where we rerank the BM25 top-100 documents with the TCT-ColBERT scores for each query-document pair.
\li \textbf{BM25>>Mono-T5} (abbreviated in Table \ref{tab:exposure-bias-results} as `\textbf{Mono-T5}') \cite{monot5}: A \textit{retrieve-and-rerank} class of model, where the top 100 documents retrieved by BM25 are subsequently reranked using a cross-encoder model MonoT5, which assigns relevance scores to each query-document pair.
\ule

\para{Evaluation Metrics}
As evaluation metrics for document exposure bias, we employ the Gini coefficient of the retrievability scores (lower scores indicating fair document exposure). As the measure for relevance, we employ the common measure of nDCG@10 \cite{ndcg10} and MAP \cite{MAP2000}.

It is important to note that, unlike per-query retrieval effectiveness measures, statistical significance testing (e.g., paired t-test) is not applicable to the Gini coefficient of retrievability or T-retrievability, since these are collection-level measures.

\begin{figure}[t]
  \centering
  \begin{subfigure}[t]{0.49\columnwidth}
    \includegraphics[width=\columnwidth]{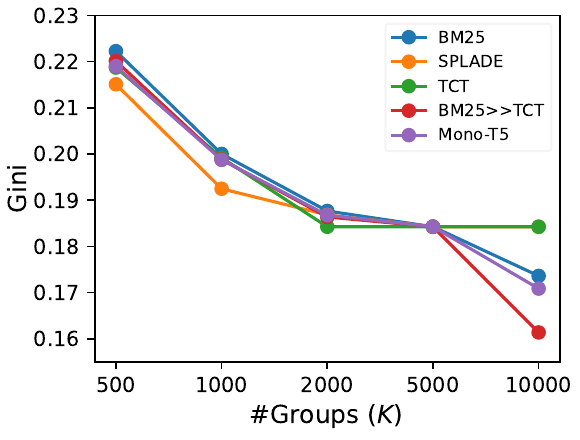}
    \caption{Min across Topics}
  \end{subfigure}
  \begin{subfigure}[t]{0.49\columnwidth}
    \includegraphics[width=\columnwidth]{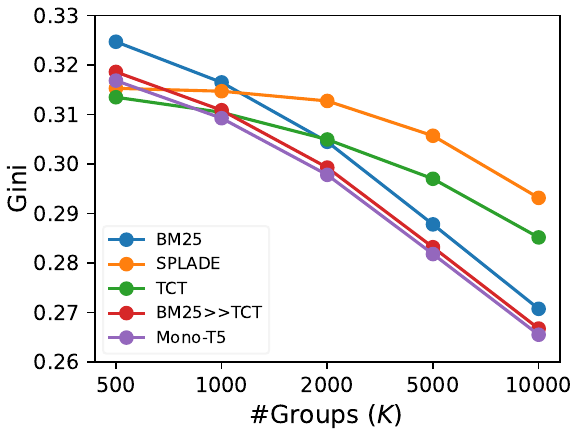}
    \caption{Mean across Topics \label{fig:aggr_gini_scikit_dense-mean}}
  \end{subfigure}
  \begin{subfigure}[t]{0.49\columnwidth}
    \centering
    \includegraphics[width=\columnwidth]{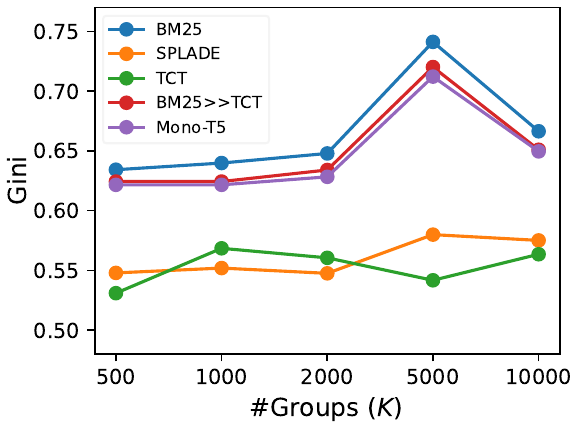}
    \caption{Max across Topics \label{fig:aggr_gini_scikit_dense-max}}
  \end{subfigure}
\caption{Variations in document exposure fairness (as measured by Gini coefficients of T-retrievability) for different granularity of topics (query groups) obtained with K-means on dense representations of the queries.}
\label{fig:aggr_gini_scikit_dense}
\end{figure}

\section{Results} \label{sec: results}

Table \ref{tab:exposure-bias-results} shows a comparison between the collection-level and the topic-focused Gini coefficients of the retrievability scores for the different IR models. In relation to RQ-1, we observe that there is considerable variance in the exposure biases across different IR models. As can be seen from the collection-level Ginis in Table \ref{tab:exposure-bias-results}, the variance ranges from the highest Gini observed for BM25 (0.4731) to the lowest for Splade (0.3948).

For RQ-2, we observe that the collection-level exposure fairness measure (Equation \ref{eq:our-ret}) is not always correlated with the more fine-grained measures (Equation \ref{eq:loc-retr}). For example, Table \ref{tab:exposure-bias-results} shows that Splade yields the lowest Gini coefficient at the collection level (0.3948), whereas Mono-T5 achieves the lowest average Gini (0.2818) when using fine-grained measures. The key observation is that our proposed fine-grained exposure fairness measures reveal further insights into a model's best, average, and worst-case behaviour.

\begin{figure}[t]
  \centering
  \begin{subfigure}[t]{0.49\columnwidth}
    \includegraphics[width=\columnwidth]{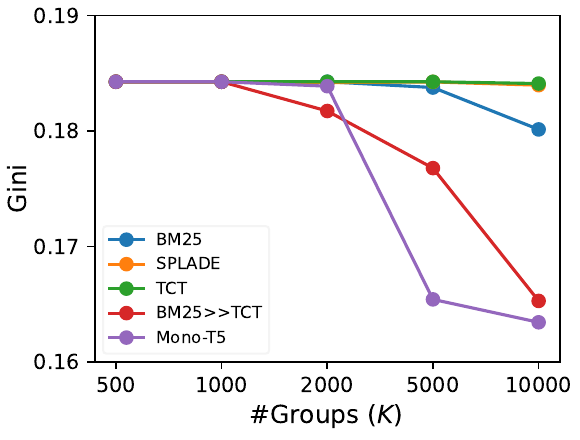}
    \caption{Min across Topics}
  \end{subfigure}
  \begin{subfigure}[t]{0.49\columnwidth}
    \includegraphics[width=\columnwidth]{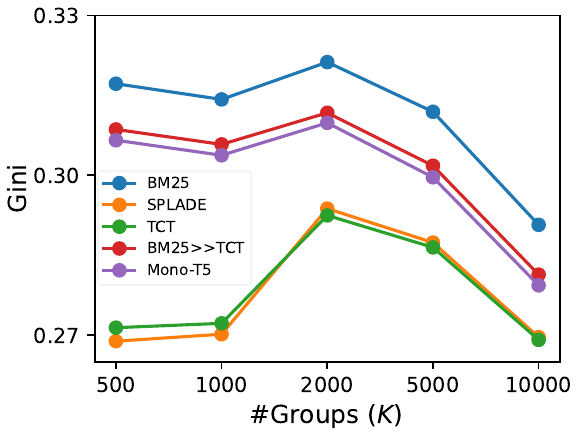}
    \caption{Mean across Topics \label{fig:aggr_gini_scikit_tfidf-mean}}
  \end{subfigure}
  \begin{subfigure}[t]{0.49\columnwidth}
    \centering
    \includegraphics[width=\columnwidth]{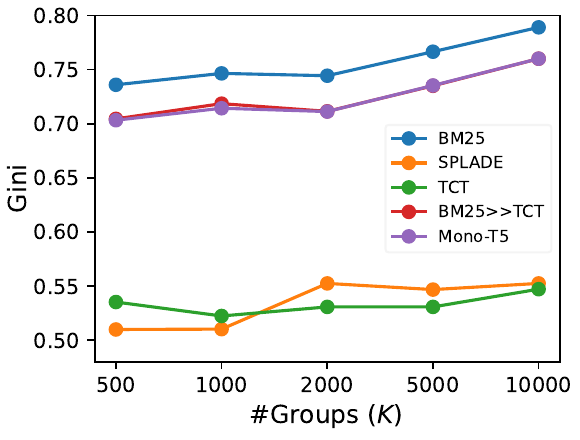}
    \caption{Max across Topics \label{fig:aggr_gini_scikit_tfidf-max}}
  \end{subfigure}

\caption{Similar to Figure \ref{fig:aggr_gini_scikit_dense} -- the only difference being that sparse (tf-idf) representation is used to cluster the queries.} 
\label{fig:aggr_gini_scikit_tfidf}
\end{figure}

In relation to RQ-3, from Figure \ref{fig:aggr_gini_scikit_dense}, we observe that both the best-case and average-case exposure fairness metrics generally decrease as the number of groups of queries increases. An exception to this trend, illustrated in Figure \ref{fig:aggr_gini_scikit_tfidf}, is that the best-case exposure fairness for Splade and TCT remains stable, and the average-case fairness for all models exhibits a slight increase in the intermediate range (i.e., with group sizes between 1000 and 2000).

As can be seen from Figure \ref{fig:aggr_gini_scikit_dense}, the topic-based retrievability analysis exhibits variations in terms of model performance for exposure fairness across a range of different topic granularities. For instance, over fine-grained topics, Splade is no longer the most fair model (as observed in Table \ref{tab:exposure-bias-results}).
Additional insights are revealed in the worst-case scenarios across both graphs, where Splade and TCT demonstrate notably higher exposure fairness (i.e., lower Gini values) as compared to other retrieval methods. The worst-case exposure fairness, mostly, tends to decrease (i.e., Gini values mostly tend to increase) when the topics become more fine-grained (i.e., as $K$ increases) thus suggesting that ranking models that appear to ensure relatively balanced document accessibility may, in fact, exhibit disparities when analysed at a finer level of query topics. 

In relation to RQ-4, it can be observed that the query grouping strategy (which in turn depends on the query representations themselves) exhibits different trends in the T-Retrievability values, e.g., the exposure bias of Splade in Figures \ref{fig:aggr_gini_scikit_tfidf-mean} and \ref{fig:aggr_gini_scikit_tfidf-max} relative to other models is much lower than those observed in Figures \ref{fig:aggr_gini_scikit_dense-mean} and \ref{fig:aggr_gini_scikit_dense-max}.

\para{Concluding Remarks}
In this paper, we analysed the exposure fairness of various retrieval models using a modified retrievability measure. We argued that measuring exposure biases by the standard retrievability-based Gini coefficient may disproportionately penalise effective ranking models, as some degree of non-uniformity can be attributed to inherent disparities in the relevance priors. Our proposed measure mitigates this effect by aggregating disparities in retrievability scores across topically focused subsets of queries. Experimental results on a standard benchmark set of a large number of queries (MS MARCO dev) indicate substantial variations in the exposure biases of different neural models.

In future, we plan to leverage a historical set of queries (e.g., the MS MARCO training set) to adjust document rankings based on prior exposure levels.

\section*{GenAI Usage Disclosure}
Generative AI tools were not used for core idea generation or experimental design. Its use was limited to minor writing and formatting.

\bibliographystyle{ACM-Reference-Format}

\balance

\bibliography{refs}

\end{document}